\documentclass[twocolumn,showkeys]{revtex4}
\usepackage[utf8]{inputenc}
\usepackage{graphicx}
\begin{document}

\title{On a hydrodynamic description of waves propagating perpendicular to the magnetic field in relativistically hot plasmas}

\author{Pavel A. Andreev}
\email{andreevpa@physics.msu.ru}
\affiliation{Department of General Physics, Faculty of physics, Lomonosov Moscow State University, Moscow, Russian Federation, 119991.}
\affiliation{Peoples Friendship University of Russia (RUDN University), 6 Miklukho-Maklaya Street, Moscow, 117198, Russian Federation.}

\date{\today}

\begin{abstract}
The novel hydrodynamic model of plasmas with the relativistic temperatures
consisted of four equations for the material fields: the concentration and the velocity field
\emph{and} the average reverse relativistic $\gamma$ functor and the flux of the reverse relativistic $\gamma$ functor
is applied to study high-frequency part of spectrum of electromagnetic waves propagating perpendicular to the external magnetic field.
The thermal effects considered for the temperatures close to the rest energy of electrons considerably change the dispersion equation
in compare with the nonrelativistic temperatures.
Analytical analysis of the changes is presented.
\end{abstract}


\maketitle





\section{Introduction}

Propagation of waves in the magnetized plasmas perpendicular to the magnetic field
with the electric field perpendicular to the external magnetic field
leads to the intermix of the longitudinal and transverse polarizations
into well known extraordinary waves.
It appears at the analysis of the plasmas in the nonrelativistic regime.
If temperatures of plasmas increases up to the relativistic temperatures
it leads to considerable changes of the hydrodynamic model and corresponding spectra of waves.
The role of relativistic temperature effects in properties of waves propagating parallel to the external field
is demonstrated in Ref. \cite{Andreev 2021 06}.
It is shown that
the circularly polarized transverse waves are considerably modified by the thermal effect
while there is no contribution of the thermal effects for the small nonrelativistic temperatures in the linear properties of these waves.
It shows that the properties of relativistic plasmas considerably differ from the properties of the nonrelativistic plasmas.
Therefore, the analysis of relativistic plasmas requires variety of models, particularly hydrodynamic models.
Here, we present the further application of the recently suggested relativistic hydrodynamic model \cite{Andreev 2021 05}.

All macroscopic models must be derived from the corresponding microscopic model.
Here we consider classic plasmas,
where particles move up to the relativistic velocities getting close to the speed of light $c$.
The concentration of particles
$n(\textbf{r},t)$ in the arbitrary inertial frame \cite{Kuz'menkov 91}, \cite{Drofa TMP 96}, \cite{Andreev PIERS 2012}
can be defined in the following form
\begin{equation}\label{RHD2021ClLM concentration definition} n(\textbf{r},t)=\frac{1}{\Delta}\int_{\Delta}d\mbox{\boldmath $\xi$}\sum_{i=1}^{N}\delta(\textbf{r}+\mbox{\boldmath $\xi$}-\textbf{r}_{i}(t)). \end{equation}
The integral operator counts the number of particles in the vicinity of the point of space.
hence, we have the number of particles in the volume $\Delta$ around point of space $\textbf{r}$ in an arbitrary moment in time $t$.

Other hydrodynamic functions are defined in the similar way \cite{Andreev 2021 05}
via operator
\begin{equation}\label{RHD2021ClLM formula for average}\langle ...\rangle\equiv\frac{1}{\Delta}\int_{\Delta}d\mbox{\boldmath $\xi$}
\sum_{i=1}^{N} ... \delta(\textbf{r}+\mbox{\boldmath $\xi$}-\textbf{r}_{i}(t)),\end{equation}
so we have
$\textbf{j}=\langle \textbf{v}_{i}(t)\rangle$,
$\textbf{v}=\textbf{j}/n$,
$\Gamma=\langle \frac{1}{\gamma_{i}}\rangle$,
$t^{a}=\langle \frac{1}{\gamma_{i}}v_{i}^{a} \rangle -\Gamma v^{a}$,
$p^{ab}=\langle v_{i}^{a}v_{i}^{b} \rangle-n v^{a}v^{b}$,
$t^{ab}=\langle \frac{1}{\gamma_{i}}v_{i}^{a}v_{i}^{b} \rangle-\Gamma v^{a}v^{b}-t^{a}v^{b}- v^{a}t^{b}$,
for $M^{abcd}$ see equation (17) of Ref. \cite{Andreev 2021 05},
here $\gamma_{i}=1/\sqrt{1-\textbf{v}_{i}(t)^{2}/c^{2}}$.
The form of these functions is not chosen, but it appears via step by step derivation of the hydrodynamic equations
as a continuous sequence.
The energy-momentum density is absence here since it is not appear in the continuity, Maxwell or other hydrodynamic equations being under consideration.
The microscopic dynamics based on the presentation of classic particles as the delta functions is considered in literature \cite{Weinberg Gr 72}.
However, it contains no explicit transition to the macroscopic scale.


This paper is organized as follows.
In Sec. II the relativistic hydrodynamic equations are presented and discussed.
In Sec. III the spectrum of collective excitations is considered analytically.
In Sec. IV a brief summary of obtained results is presented.


\section{Relativistic hydrodynamic model}

Here we follow Refs. \cite{Andreev 2021 06}, \cite{Andreev 2021 05}
where a set of relativistic hydrodynamic equations
is obtained and applied for plasmas with the relativistic temperatures.
The model is composed of four equations.
There are other hydrodynamic models of high-temperature relativistic plasmas,
where the interaction of particles is considered in terms of the momentum evolution equation
\cite{Hazeltine APJ 2002}, \cite{Shatashvili ASS 97}, \cite{Shatashvili PoP 99}, \cite{Shatashvili PoP 20}, \cite{Mahajan PoP 2002}.

First equation is the continuity equation
\begin{equation}\label{RHD2021ClLM cont via v} \partial_{t}n+\nabla\cdot(n\textbf{v})=0.\end{equation}

Next, the velocity field evolution equation is
$$n\partial_{t}v^{a}+n(\textbf{v}\cdot\nabla)v^{a}+\frac{\partial^{a}p}{m}
=\frac{e}{m}\Gamma E^{a}+\frac{e}{mc}\varepsilon^{abc}(\Gamma v^{b}+t^{b})B^{c}$$
\begin{equation}\label{RHD2021ClLM Euler for v} -\frac{e}{mc^{2}}(\Gamma v^{a} v^{b}+v^{a}t^{b}+v^{b}t^{a})E^{b}
-\frac{e}{mc^{2}}\tilde{t}E^{a}, \end{equation}
where $p$ is the flux of the thermal velocities.

The equation of evolution of the averaged reverse relativistic gamma factor is
\begin{equation}\label{RHD2021ClLM eq for Gamma} \partial_{t}\Gamma+\partial_{b}(\Gamma v^{b}+t^{b})
=-\frac{e}{mc^{2}}n\textbf{v}\cdot\textbf{E}\biggl(1-\frac{1}{c^{2}}\biggl(\textbf{v}^{2}+\frac{5p}{n}\biggr)\biggr).\end{equation}
Function $\Gamma$ is also called the hydrodynamic Gamma function \cite{Andreev 2021 05}.

The final equation in this set of hydrodynamic equations is the equation of evolution for the thermal part of
current of the reverse relativistic gamma factor (the hydrodynamic Theta function):
$$(\partial_{t}+\textbf{v}\cdot\nabla)t^{a}+\partial_{a}\tilde{t}
+(\textbf{t}\cdot\nabla) v^{a}+t^{a} (\nabla\cdot \textbf{v})$$
$$+\Gamma(\partial_{t}+\textbf{v}\cdot\nabla)v^{a}
=\frac{e}{m}nE^{a}\biggl[1-\frac{\textbf{v}^{2}}{c^{2}}-\frac{3p}{nc^{2}}\biggr]$$
$$+\frac{e}{mc}\varepsilon^{abc}nv^{b}B^{c}\biggl[1-\frac{\textbf{v}^{2}}{c^{2}}-\frac{5p}{nc^{2}}\biggr]
-\frac{2e}{mc^{2}}E^{a}p\biggl[1-\frac{\textbf{v}^{2}}{c^{2}}\biggr]$$
\begin{equation}\label{RHD2021ClLM eq for t a} -\frac{e}{mc^{2}}nv^{a}v^{b}E^{b}\biggl[1-\frac{\textbf{v}^{2}}{c^{2}}-\frac{9p}{nc^{2}}\biggr]
-\frac{10e}{3mc^{4}}M E^{a}.\end{equation}
Function $M$ is presented via the equation of state.
All hydrodynamic equations are obtained in the mean-field approximation (the self-consistent field approximation).

The equations of electromagnetic field have the traditional form
presented in the three-dimensional notations
$ \nabla \cdot\textbf{B}=0$, $\nabla \cdot\textbf{E}=4\pi(en_{i}-en_{e})$,
\begin{equation}\label{RHD2021ClLM rot E and div E}
\nabla\times \textbf{E}=-\frac{1}{c}\partial_{t}\textbf{B},
\end{equation}
and
\begin{equation}\label{RHD2021ClLM rot B with time}
\nabla\times \textbf{B}=\frac{1}{c}\partial_{t}\textbf{E}+\frac{4\pi q_{e}}{c}n_{e}\textbf{v}_{e},\end{equation}
where the ions exist as the motionless background.


\section{Extraordinary waves in the relativistic magnetized plasmas}

We consider small amplitude collective excitations of the macroscopically motionless equilibrium state.
This equilibrium state is described by the relativistic Maxwellian distribution.
The equilibrium state is described within equilibrium concentration $n_{0}$.
The velocity field $\textbf{v}_{0}$ in the equilibrium state is equal to zero.
The equilibrium electric field $\textbf{E}_{0}$ is also equal to zero.
The plasma is located in the constant and uniform external magnetic field $\textbf{B}_{0}=B_{0}\textbf{e}_{z}$.
Two second rank tensors and one fourth rank tensor are involved in the description of the thermal effects.
The symmetric tensors $p^{ab}$ and $t^{ab}$ are assumed to be diagonal tensors:
$p^{ab}=p\delta^{ab}$ and $t^{ab}=\tilde{t}\delta^{ab}$.
The "diagonal" form is assumed for the symmetric fourth rank tensor $M^{abcd}$ as well:
$M^{abcd}=M_{0}(\delta^{ab}\delta^{cd}+\delta^{ac}\delta^{bd}+\delta^{ad}\delta^{bc})/3$.
We consider propagation of perturbations in the direction perpendicular to the external magnetic field $\textbf{k}=\{ k_{x},0,0\}$.

The following equilibrium functions
$\Gamma_{0}$, $\textbf{t}_{0}$, $p_{0}$, $\tilde{t}_{0}$, $\textbf{q}_{0}$, $M_{0}$
are involved in the description of the equilibrium state
$p^{ab}=c^{2}\delta^{ab}\tilde{Z}f_{1}(\beta)/3$,
$t^{ab}=c^{2}\delta^{ab}\tilde{Z}f_{2}(\beta)/3$
$M^{abcd}=c^{4}(\delta^{ab}\delta^{cd}+\delta^{ac}\delta^{bd}+\delta^{ad}\delta^{bc})\tilde{Z}f_{3}(\beta)/15$,
$\Gamma_{0}=n_{0}K_{1}(\beta)/K_{2}(\beta)$,
and
$\textbf{q}=0$,
where
$\beta=mc^{2}/T$,
$\tilde{Z}=4\pi Z (mc)^{3}=n\beta K_{2}^{-1}(\beta)$,
\begin{equation}\label{RHD2021ClLM f 1} f_{1}(\beta)=\int_{1}^{+\infty}\frac{d x}{x}(x^{2}-1)^{3/2}e^{-\beta x}, \end{equation}
\begin{equation}\label{RHD2021ClLM f 2} f_{2}(\beta)=\int_{1}^{+\infty}\frac{d x}{x^{2}}(x^{2}-1)^{3/2}e^{-\beta x}, \end{equation}
and
\begin{equation}\label{RHD2021ClLM f 3} f_{3}(\beta)=\int_{1}^{+\infty}\frac{d x}{x^{3}}(x^{2}-1)^{5/2}e^{-\beta x}. \end{equation}
Functions $f_{1}(\beta)$, $f_{2}(\beta)$ and $f_{3}(\beta)$ are calculated numerically below for the chosen values of temperatures.
We introduce three characteristic velocities
$\delta p=U_{p}^{2} \delta n$,
$\delta \tilde{t}=U_{t}^{2} \delta n$,
and
$\delta M=U_{M}^{4} \delta n$.

Let us present the linearized equations for the plane wave excitations.
For instance for the concentration we have $\delta n=N_{0}e^{-\imath\omega t+\imath k_{x}x}$,
where $\omega$ is the frequency, $N_{0}$ is the constant amplitude.
We start with the linearized continuity equation (\ref{RHD2021ClLM cont via v}):
\begin{equation}\label{RHD2021ClLM continuity equation lin 1D}
-\imath\omega\delta n+n_{0}\imath k_{x} \delta v_{x}=0. \end{equation}
Next, we show the linearized equations for the evolution of the three projections of velocity field
\begin{equation}\label{RHD2021ClLM velocity field evolution equation lin 1D x}
-\imath\omega n_{0}\delta v_{x} +\imath k_{x}\frac{\delta p}{m}
=\frac{q_{e}}{m}\biggl(\Gamma_{0} -\frac{\tilde{t}_{0}}{c^{2}}\biggr)\delta E_{x} +\Omega_{e}(\Gamma_{0}\delta v_{y}+\delta t_{y}),
\end{equation}
\begin{equation}\label{RHD2021ClLM velocity field evolution equation lin 1D y}
-\imath\omega n_{0}\delta v_{y}
=\frac{q_{e}}{m}\biggl(\Gamma_{0} -\frac{\tilde{t}_{0}}{c^{2}}\biggr)\delta E_{y}-\Omega_{e}(\Gamma_{0}\delta v_{x}+\delta t_{x}),
\end{equation}
and
\begin{equation}\label{RHD2021ClLM velocity field evolution equation lin 1D z}
-\imath\omega n_{0}\delta v_{z}
=\frac{q_{e}}{m}\Gamma_{0} \delta E_{z}-\frac{q_{e}}{mc^{2}}\tilde{t}_{0}\delta E_{z},
\end{equation}
where
$\Omega_{e}=q_{e}B_{0}/mc$ is the cyclotron frequency.

The Maxwell equations lead to
\begin{equation}\label{RHD2021ClLM Maxwell lin wave x}
\omega^{2}\delta E_{x}+4\pi q_{e}\imath\omega n_{0}\delta v_{x}=0, \end{equation}
\begin{equation}\label{RHD2021ClLM Maxwell lin wave y}
(\omega^{2}-k_{x}^{2}c^{2})\delta E_{y}+4\pi q_{e}\imath\omega n_{0}\delta v_{y}=0, \end{equation}
and
\begin{equation}\label{RHD2021ClLM Maxwell lin wave z}
(\omega^{2}-k_{x}^{2}c^{2})\delta E_{z}+4\pi q_{e}\imath\omega n_{0}\delta v_{z}=0. \end{equation}


To get a closed set of equations including the x- and y-projections of the velocity field and the x- and y-projections of the electric field
we need to include the evolution equations for the x- and y-projections of the flux of the reverse relativistic factor:
$$-\imath\omega \delta t_{x} -\imath\omega\Gamma_{0} \delta v_{x} +\imath k_{x}\delta \tilde{t}$$
\begin{equation}\label{RHD2021ClLM evolution of Theta lin 1D x}
=n_{0}\biggl(1-\frac{5 p_{0}}{n_{0}c^{2}}\biggr)\biggl[\frac{q_{e}}{m}\delta E_{x}+\Omega_{e}\delta v_{y}\biggr] -\frac{10q_{e}}{3mc^{4}}M_{0}\delta E_{x}, \end{equation}
and
$$-\imath\omega\delta t_{y} -\imath\omega\Gamma_{0} \delta v_{y}$$
\begin{equation}\label{RHD2021ClLM evolution of Theta lin 1D y}
=n_{0}\biggl(1-\frac{5 p_{0}}{n_{0}c^{2}}\biggr)\biggl[\frac{q_{e}}{m}\delta E_{y}-\Omega_{e}\delta v_{x}\biggr] -\frac{10q_{e}}{3mc^{4}}M_{0}\delta E_{y}. \end{equation}
where $M_{0}^{xxcc}=(5/3)M_{0}$.


Let us start the discussion of small amplitude waves in the liner approximation with
the presentation the spectrum in nonrelativistic regime:
\begin{equation}\label{RHD2021ClLM Spectrum nonRel}
\omega^{2}-k^{2}c^{2}-\omega_{Le}^{2}\frac{2\omega^{2}-k_{x}^{2}u^{2}-k_{x}^{2}c^{2}-\omega_{Le}^{2}}{\omega^{2}-k_{x}^{2}u^{2}-\Omega_{e}^{2}}=0,
\end{equation}
where
$\omega_{Le}^{2}=4\pi e^{2}n_{0}/m$ is the Langmuir frequency,
$\Omega_{e}=-\mid \Omega_{e}\mid=q_{e}B_{0}/mc$.
Equation (\ref{RHD2021ClLM Spectrum nonRel})
is found from the standard hydrodynamic model based on the continuity and Euler equations.
Equation (\ref{RHD2021ClLM Spectrum nonRel})
is found from the following equality to zero of the determinant of matrix appearing from the equations for the projections of the electric field
\begin{widetext}
\begin{equation}\label{RHD2021ClLM Dispersion determinant nonRel}
\hat{\varepsilon}_{NR}=\left(
                   \begin{array}{cc}
                     \varepsilon_{xx,NR} & \varepsilon_{xy,NR} \\
                     \varepsilon_{yx,NR} & \varepsilon_{yy,NR} \\
                   \end{array}
                 \right)=
\left(
  \begin{array}{cc}
    -1+\frac{\omega_{Le}^{2}}{\omega^{2}-k_{x}^{2}u^{2}-\Omega_{e}^{2}}, & \imath\frac{\Omega_{e}}{\omega}\frac{\omega_{Le}^{2}}{\omega^{2}-k_{x}^{2}u^{2}-\Omega_{e}^{2}} \\
    -\imath\frac{\Omega_{e}}{\omega} \frac{\omega_{Le}^{2}}{\omega^{2}-\Omega_{e}^{2}}\biggl[1+\frac{k_{x}^{2}u^{2}}{\omega^{2}-k_{x}^{2}u^{2}-\Omega_{e}^{2}}\biggr], & \frac{k_{x}^{2}c^{2}}{\omega^{2}}-1 +\frac{\omega_{Le}^{2}}{\omega^{2}-\Omega_{e}^{2}}\biggl[1+\frac{k_{x}^{2}u^{2}}{\omega^{2}-k_{x}^{2}u^{2}-\Omega_{e}^{2}}\biggr] \\
  \end{array}
\right).
\end{equation}
Let us to point out that this matrix is not symmetric.
Moreover, it is not Hermitian.

Let us present corresponding matrix for the relativistically hot plasmas obtained from the linearized equations presented above:
\begin{equation}\label{RHD2021ClLM Dispersion determinant}
\hat{\varepsilon}=\left(
  \begin{array}{cc}
    -1+\omega_{Le}^{2}\Sigma, & \omega_{Le}^{2} \Xi\\
    \frac{\omega_{Le}^{2}}{\omega^{2}-\Omega_{e}^{2}(1-\frac{5u_{p}^{2}}{c^{2}})}
    \biggl[\biggl(-\imath\frac{\Omega_{e}}{\omega}(1-\frac{5u_{p}^{2}}{c^{2}})-\frac{10u_{M}^{4}}{3c^{4}}\biggr)
    -\imath\frac{\Omega_{e}}{\omega}k_{x}^{2}u_{t}^{2} \Sigma\biggr],
    &
\frac{k_{x}^{2}c^{2}}{\omega^{2}}-1
+\frac{\omega_{Le}^{2}}{\omega^{2}-\Omega_{e}^{2}(1-\frac{5u_{p}^{2}}{c^{2}})} \biggl[\frac{\Gamma_{0}}{n_{0}}-\frac{u_{t}^{2}}{c^{2}} +k_{x}^{2}u_{t}^{2}(-\imath\frac{\Omega_{e}}{\omega})\Xi\biggr] \\
  \end{array}
\right),
\end{equation}\end{widetext}
where
\begin{equation}\label{RHD2021ClLM Sigma}
\Sigma\equiv\frac{(\frac{\Gamma_{0}}{n_{0}}-\frac{u_{t}^{2}}{c^{2}})}{\omega^{2}-k_{x}^{2}u_{p}^{2}-\Omega_{e}^{2}(1-\frac{5u_{p}^{2}}{c^{2}})},
\end{equation}
and
\begin{equation}\label{RHD2021ClLM Xi}
\Xi\equiv \frac{(\imath\frac{\Omega_{e}}{\omega} (1-\frac{5u_{p}^{2}}{c^{2}})-\frac{10u_{M}^{4}}{3c^{4}})}{\omega^{2}-k_{x}^{2}u_{p}^{2}-\Omega_{e}^{2}(1-\frac{5u_{p}^{2}}{c^{2}})}.
\end{equation}

Characteristic thermal velocities $u_{p}$ and $u_{t}$ reduce to the traditional nonrelativistic thermal velocity $u$ appearing from the pressure
(see equations (\ref{RHD2021ClLM Spectrum nonRel}) and (\ref{RHD2021ClLM Dispersion determinant nonRel})).

Element $\varepsilon_{xx,NR}$ at the transition to the relativistic case (\ref{RHD2021ClLM Dispersion determinant}) has two modifications.
First, we have $\omega_{Le}^{2}(\Gamma_{0}/n_{0}-u_{t}^{2}/c^{2})$ instead of $\omega_{Le}^{2}$ existing in the nonrelativistic regime.
Such modification of the Langmuir frequency is not universal.
In other elements of the matrix we have different representation of the Langmuir frequency.
Second, the square of the cyclotron frequency $\Omega_{e}^{2}$ being in the denumerator is replaced by the thermally modified version
$\Omega_{e}^{2}(1-5u_{p}^{2}/c^{2})$.
It leads to the effective decrease of the cyclotron frequency by the thermal motion.
This modification of the cyclotron frequency square $\Omega_{e}^{2}$ being in the denumerators is correct for all four elements of the tensor.
But the cyclotron frequency located in the numerator has different transformation.

Element $\varepsilon_{xy,NR}$ transforms in two ways.
First, the square of the cyclotron frequency $\Omega_{e}^{2}$ being in the denumerator is replaced by the thermally modified version
$\Omega_{e}^{2}(1-5u_{p}^{2}/c^{2})$.
Second transformation is related to the representation of the cyclotron frequency $\Omega_{e}$ in the numerator,
where
$\imath\frac{\Omega_{e}}{\omega}$
transforms into
$(\imath\frac{\Omega_{e}}{\omega} (1-\frac{5u_{p}^{2}}{c^{2}})-\frac{10u_{M}^{4}}{3c^{4}})$.
We see that combination $\imath\frac{\Omega_{e}}{\omega}$ multiplies by $(1-\frac{5u_{p}^{2}}{c^{2}})$,
but the additional term proportional to $u_{M}$ appears.

In the nonrelativistic limit element $\varepsilon_{yx,NR}$ has nontrivial difference in compare with element $\varepsilon_{xy,NR}$.
The difference between elements $\varepsilon_{yx}$ and $\varepsilon_{xy}$ is more pronunced.
Element $\varepsilon_{yx,NR}$ is proportional to the sum of two terms
$-\imath\frac{\Omega_{e}}{\omega}$
and
$-\imath\frac{\Omega_{e}}{\omega}\frac{k_{x}^{2}u^{2}}{\omega^{2}-k_{x}^{2}u^{2}-\Omega_{e}^{2}}$.
The first of them $-\imath\frac{\Omega_{e}}{\omega}$ represents in the way similar to element $\varepsilon_{xy,NR}$:
$-\imath\frac{\Omega_{e}}{\omega} \rightarrow -\imath\frac{\Omega_{e}}{\omega}(1-\frac{5u_{p}^{2}}{c^{2}})-\frac{10u_{M}^{4}}{3c^{4}}$.
The second of them $-\imath\frac{\Omega_{e}}{\omega}\frac{k_{x}^{2}u^{2}}{\omega^{2}-k_{x}^{2}u^{2}-\Omega_{e}^{2}}$
is proportional to $k_{x}^{2}u^{2}$.
While it has following transformation
$k_{x}^{2}u^{2}\rightarrow k_{x}^{2}u_{t}^{2}(\Gamma_{0}/n_{0}-u_{t}^{2}/c^{2})$.
Moreover, the square of the cyclotron frequency $\Omega_{e}^{2}$ being in the denumerator transforms in the way described above
$\Omega_{e}^{2}\rightarrow\Omega_{e}^{2}(1-5u_{p}^{2}/c^{2})$.


Element $\varepsilon_{yy,NR}$ composed of the electromagnetic part
$\frac{k_{x}^{2}c^{2}}{\omega^{2}}-1$,
which does not change,
and the matter part
$\frac{\omega_{Le}^{2}}{\omega^{2}-\Omega_{e}^{2}}\biggl[1+\frac{k_{x}^{2}u^{2}}{\omega^{2}-k_{x}^{2}u^{2}-\Omega_{e}^{2}}\biggr]$,
which has modifications caused by the thermal relativistic effects.
In two parts of this equation we see the square of the cyclotron frequency $\Omega_{e}^{2}$
which is transformed in the same way as ir changed in other elements of matrix
$\Omega_{e}^{2}\rightarrow\Omega_{e}^{2}(1-5u_{p}^{2}/c^{2})$.
Further description of the modification of the matter part can be described at the splitting of this part on two terms:
$\omega_{Le}^{2}$ and
$\frac{\omega_{Le}^{2}\cdot k_{x}^{2}u^{2}}{\omega^{2}-k_{x}^{2}u^{2}-\Omega_{e}^{2}}$.
The singular square of the Langmuir frequency mofdifies in the way similar to element $\varepsilon_{xx}$:
$\omega_{Le}^{2}\rightarrow \omega_{Le}^{2}(\Gamma_{0}/n_{0}-u_{t}^{2}/c^{2})$.
The numerator of the second term $\omega_{Le}^{2}\cdot k_{x}^{2}u^{2}$ transforms into
$\omega_{Le}^{2}\cdot k_{x}^{2}u_{t}^{2}\cdot (-\imath\frac{\Omega_{e}}{\omega})$
$\cdot [\imath\frac{\Omega_{e}}{\omega} (1-\frac{5u_{p}^{2}}{c^{2}})-\frac{10u_{M}^{4}}{3c^{4}}]$.

Evolution of $\delta E_{z}$ leads to the spectrum of the ordinary electromagnetic wave
with the linear transverse polarization of the electromagnetic field:
\begin{equation}\label{RHD2021ClLM ordinary}
\omega^{2}=k_{x}^{2}c^{2}+\omega_{Le}^{2}\biggl(\frac{\Gamma_{0}}{n_{0}}-\frac{u_{t}^{2}}{c^{2}}\biggr),
\end{equation}
where the Langmuir frequency is thermally modified in the same way as in element $\varepsilon_{xx}$ in equation (\ref{RHD2021ClLM Dispersion determinant})
or as in the spectrum of the Langmuir waves \cite{Andreev 2021 05}.

\section{Conclusion}

Contribution of the thermal-relativistic effects in the spectrum of the extraordinary waves in the magnetized plasmas
has been illustrated analytically.
To this end the explicit form of the dielectric permeability tensor has been found.
To illustrate the relativistic-thermal effects this tensor is compared with the corresponding nonrelativistic tensor.

Considerable modification of the dielectric permeability tensor has been obtained.
It demonstrates the background for the further analysis of the linear and nonlinear wave phenomena in the relativistically hot plasmas
within the presented hydrodynamic model based on the dynamics of
four material fields: the concentration and the velocity field
\emph{and} the average reverse relativistic $\gamma$ functor and the flux of the reverse relativistic $\gamma$ functor.

\section{Acknowledgements}

Work is supported by the Russian Foundation for Basic Research (grant no. 20-02-00476).
This paper has been supported by the RUDN University Strategic Academic Leadership Program.

\section{DATA AVAILABILITY}

Data sharing is not applicable to this article as no new data were
created or analyzed in this study, which is a purely theoretical one.

\appendix

\section{Appendix:Numerical estimation of the relativistic-thermal parameters}

Three temperature regimes are chosen $T_{1}=0.1 mc^{2}$, $T_{2}=mc^{2}$, and $T_{3}=10mc^{2}$.
It gives the following values of the dimensionless reverse temperature $\beta=mc^{2}/T$:
$\beta_{1}=10$, $\beta_{2}=1$, and $\beta_{3}=0.1$.

For the relatively small relativistic temperature $\beta_{1}=10$,
we calculate $K_{1}/K_{2}=0.91$, $U_{t}^{2}/c^{2}=0.07$,
$U_{p}^{2}/c^{2}=0.08$, and $U_{M}^{4}/c^{4}=0.02$,
where
$K_{1}(10)=2\times 10^{-5}$, $K_{2}(10)=2.2\times10^{-5}$, $f_{1}(10)=5\times 10^{-7}$,
$f_{2}(10)=4.2\times10^{-7}$, $f_{3}(10)=1.7\times 10^{-7}$.

For $\beta_{2}=1$, we get $K_{1}/K_{2}=0.38$, $U_{t}^{2}/c^{2}=\beta f_{2}/(3 K_{2})=0.1$,
$U_{p}^{2}/c^{2}=\beta f_{1}/(3 K_{2})=0.28$, and $U_{M}^{4}/c^{4}=\beta f_{3}/(5 K_{2})=0.15$,
where $K_{1}(1)=0.6$, $K_{2}(1)=1.6$, $f_{1}(1)=1.35$, $f_{2}(1)=0.46$, $f_{3}(1)=1.17$.

For $\beta_{3}=0.1$, we have $K_{1}/K_{2}=0.05$, $U_{t}^{2}/c^{2}=\beta f_{2}/(3 K_{2})=0.02$,
$U_{p}^{2}/c^{2}=0.33$, and $U_{M}^{4}/c^{4}=0.2$,
where $K_{1}(0.1)=10$, $K_{2}(0.1)=200$, $f_{1}(0.1)=2\times 10^{3}$, $f_{2}(0.1)=100$, $f_{3}(0.1)=2\times 10^{3}$.

\end{document}